\documentclass[fleqn,usenatbib]{mnras}

\usepackage{newtxtext,newtxmath}

\usepackage[T1]{fontenc}

\DeclareRobustCommand{\VAN}[3]{#2}
\let\VANthebibliography\thebibliography
\def\thebibliography{\DeclareRobustCommand{\VAN}[3]{##3}\VANthebibliography}

\usepackage{graphicx}	
\usepackage{amsmath}	
\usepackage[dvipsnames]{xcolor}

\newcommand{\msun}{$M_{\odot}$}

\newcommand{\gasdust}{$\Delta_\text{g/d}$}
\newcommand{\rsub}{$R_\text{sub}$}
\newcommand{\rgap}{$R_\text{gap}$}
\newcommand{\rcav}{$R_\text{cav}$}

\title[SIMBA Chemical Solver]{\resizebox{\textwidth}{!}{SIMBA: A Python-based single-point astrochemical solver and analysis tool}}

\author[Keyte \& Ran]{
Luke Keyte,$^{1}$\thanks{E-mail: l.keyte@qmul.ac.uk}
Jason Ran$^{2}$
\\
$^{1}$Astronomy Unit, School of Physics and Astronomy, Queen Mary University of London, London E1 4NS, United Kingdom.\\
$^{2}$Department of Physics and Astronomy, University College London, Gower Street, WC1E 6BT London, United Kingdom.
}

\date{Accepted 2025 September 15. Received 2025 September 10; in original form 2025 July 11.}

\pubyear{\the\year{}}

\begin{document}
\label{firstpage}
\pagerange{\pageref{firstpage}--\pageref{lastpage}}
\maketitle

\begin{abstract}
Determining molecular abundances in astrophysical environments is crucial for interpreting observational data and constraining physical conditions in these regions. Chemical modelling tools are essential for simulating the complex processes that govern molecular evolution. We present \textsc{simba}, a new Python-based single-point astrochemical modelling package designed to solve chemical reaction networks across diverse astrophysical environments. The software follows standardised rate equation approaches to evolve molecular abundances under specified physical conditions, incorporating gas-phase chemistry, grain-surface processes, and photochemistry. While leveraging Python for accessibility, performance-critical routines utilise just-in-time compilation to achieve computational efficiency suitable for research applications. A key feature of \textsc{simba} is its graphical interface, which enables rapid investigation of chemical evolution under varying physical conditions. This makes it particularly valuable for exploring parameter dependencies and complementing more computationally intensive multi-dimensional models. We demonstrate the package's capabilities by modelling chemical evolution in a photoevaporative flow driven by external FUV irradiation. Using simplified gas dynamics, we chain multiple \textsc{simba} instances to create a dynamic 1D model where gas evolves both chemically and dynamically. Comparing this approach to typical `static' models—where chemistry in each grid cell evolves independently—reveals that molecular ices, especially those with relatively high binding energies like H$_2$O, can survive much farther into the flow than static models predict. This example case highlights how \textsc{simba} can be extended to higher dimensions for investigating complex chemical processes. The package is open-source and includes comprehensive documentation.
\end{abstract}

\begin{keywords}
astrochemistry -- methods: numerical --methods: data analysis -- ISM: abundances -- protoplanetary discs
\end{keywords}



\section{Introduction}

Chemical models are essential tools in modern astrophysics, enabling the interpretation of observational data such as molecular line fluxes and spectra. These models simulate the complex chemistry occurring across diverse astrophysical environments, including the interstellar medium (ISM), molecular clouds, photodissociation regions (PDRs), and protoplanetary disks. Through systematic simulation of chemical processes under varying conditions, they provide crucial insights into the underlying physical and chemical properties of these regions.

The foundation of astrochemical computational modelling lies in the single-point (0D) chemical solver, which computes molecular abundances for a given set of physical conditions. These solvers evolve a chemical network by solving rate equations, taking fixed parameters such as dust and gas densities, temperatures, and radiation fields as inputs. The network comprises a defined set of atomic and molecular species with their initial abundances, alongside chemical reactions and associated rate parameters, typically sourced from established databases such as UMIST \citep{umist_2022} and KIDA \citep{kida_2012}. Solutions are obtained either through time integration of the rate equations or by solving for steady-state conditions.

This basic framework can be extended to more complex environments by implementing multi-dimensional grids (1D, 2D, or 3D) where each cell represents an independent chemical calculation based on local conditions. For example, chemical models of protoplanetary disks often employ 2D grids in radius and height, with density and temperature structures derived from analytical prescriptions or separate physical models \citep[e.g.][]{cleeves_2016, kama_2016b, van_der_marel_2016, fedele_2017, miotello_2017, maps_5_zhang_2021, maps_7_bosman_2021, keyte_2024a, keyte_2024b, leemker_2024}. While most multi-dimensional implementations maintain static physical conditions, advanced implementations can incorporate time-varying structures to simulate dynamic processes such as molecular cloud collapse \citep[e.g.][]{aikawa_2012}, pebble growth in protoplanetary disks \citep{van_clepper_2022}, or supernovae ejecta evolution \citep{cherchneff_dwerk_2009}.

A number of established astrochemical codes have been developed to support these efforts, each with unique strengths. Codes such as \textsc{krome} \citep{krome_grassi_2014}, \textsc{nautilus} \citep{nautilus_ruaud_2016}, \textsc{alchemic} \citep{alchemic_semenov_2010}, \textsc{astrochem} \citep{astrochem_maret_2015}, and \textsc{uclchem} \citep{uclchem_holdship_2017} enable detailed gas-grain chemistry calculations across diverse environments including diffuse/dense clouds, PDRs, pre-stellar cores, and protostars \citep[e.g.][]{van_borm_2014, moorkerjea_2016, koumpia_2017, legal_2017, manna_pal_2024}. Specialised codes like \textsc{dali} \citep{bruderer_2012, bruderer_2013} and \textsc{prodimo} \citep{prodimo_woitke_2009} offer comprehensive frameworks specifically optimised for protoplanetary disk chemistry. Many of these codes integrate chemical solvers into broader frameworks, incorporating additional physics such as Monte Carlo radiative transfer, gas heating/cooling processes, and dust dynamics. These established codes, typically written in high-performance languages like Fortran or C, offer robust computational efficiency crucial for handling complex chemical networks and high-resolution grids. However, this optimisation often comes at the cost of accessibility, with steep learning curves that can present barriers to newcomers in the field.

Meanwhile, the Python programming language has emerged as the primary tool in astronomy for data analysis, visualisation, and pipeline development, valued for its accessibility, readability, and extensive ecosystem of scientific libraries \citep[e.g.][]{matplotlib_hunter_2007, scikitlearn_2011, numba_lam_2015, astropy_2018, scipy_2020}. While traditionally less computationally efficient than languages like C or Fortran for intensive simulations, Python's emphasis on code clarity and ease-of-use makes it particularly well-suited for developing accessible scientific tools. In the context of astrochemistry, the main Python-based single-point chemical solver developed to date is \textsc{ggchempy}, which is primarily designed for modelling chemistry in the ISM and dense cloud cores \citep{ggchempy_ge_2022}.

Building on these foundations, we present \textsc{simba} (`Solver for Inferring Molecular aBundances in Astrophysical environments'), a new chemical solver and analysis tool implemented entirely in Python. \textsc{simba} extends the capabilities of existing Python-based tools by enabling chemistry calculations across more diverse environments and incorporating additional processes. While prioritizing accessibility and understanding over raw computational speed, the code employs \texttt{Numba} \citep{numba_lam_2015} optimisation to achieve performance suitable for most research applications. \textsc{simba} also features a modern graphical interface (GUI) for results analysis and visualisation, and tools for exploring abundances, reaction rates, and chemical pathways. This combination of features makes it particularly valuable for researchers new to the field and as an educational tool, while maintaining the scientific rigour required for research applications. The Python implementation also allows for straightforward integration into existing workflows and enables users to easily modify and extend the code's capabilities.

This paper is structured as follows: Section 2 presents a detailed description of the chemical model and code implementation. Section 3 provides benchmark comparisons against the established thermochemical code \textsc{DALI}. In Section 4, we demonstrate \textsc{simba}'s capabilities through an example application: modelling the chemical evolution of material in an externally-driven photoevaporative flow.

\section{Chemical Model and Code Implementation}
\label{sec:model_and_code}

\subsection{Chemical model description}
\label{subsec:chemical_model}

The chemical model in \textsc{simba} follows the standard rate equation approach to evolve molecular abundances under specified physical conditions. The model solves a system of coupled ordinary differential equations (ODEs) that describe the time evolution of species number densities:
\begin{equation}
\frac{dn_i}{dt} = -\,n_i \sum_j k_{ij} n_j + \sum_j n_j \sum_l k_{jl} n_l ,
\end{equation}
where $n_i(t)$ is the number density (cm$^{-3}$) of species $i$ at time $t$. The first term on the right-hand side represents the destruction of species $i$ through two-body reactions with partners $j$, with rate coefficients $k_{ij}$. The second term represents the formation of species $i$ from two-body reactions in which a reactant $j$ combines with another species $l$, governed by coefficients $k_{jl}$. This expression explicitly shows only two-body reactions; in practice, additional terms can be included to represent first-order processes or higher-order reactions.

The chemical network includes gas-phase reactions, grain-surface chemistry, and photochemical processes. Gas-phase reactions follow standard temperature-dependent rate coefficients of the modified Arrhenius form:
\begin{equation}
    k(T) = \alpha \left(\frac{T_\text{gas}}{300\,\text{K}}\right)^\beta \exp\left(-\frac{\gamma}{T_\text{gas}}\right),
\end{equation}
where $\alpha$, $\beta$, and $\gamma$ are reaction-specific parameters, and $T_\text{gas}$ is the gas temperature. Each reaction has defined temperature limits ($T_\text{min}$ and $T_\text{max}$) that constrain its validity range. While alternative expressions for gas-phase reactions exist \citep[e.g.][]{galli_palla_1998, glover_jappsen_2007}, these are not natively included in the current model. However, \textsc{simba}’s modular structure means such reaction types can be readily added by users if desired (see Section \ref{subsubsec:code_overview}).

Photochemical processes, including direct photodissociation and photoionization, are parametrised using the form:
\begin{equation}
    k_\text{ph} = \alpha G_0 \exp(-\gamma A_\text{V}),
\end{equation}
where $G_0$ is the FUV field strength in Draine units ($\sim 2.7 \times 10^{-3}$ erg s$^{-3}$ cm$^{-2}$, integrated between 911-2067 \AA), $A_\text{V}$ is the visual extinction, and $\alpha$ and $\gamma$ are reaction-specific parameters. The model can optionally account for self-shielding effects in CO, N$_2$, H$_2$, and atomic C, following the prescriptions of \citet{visser_2009},\citet{visser_2018}, \citet{draine_bertoldi_1996}, and \citet{kamp_bertoldi_2000}.

To calculate these self-shielding factors, the code requires the total hydrogen column density ($N_\text{H}$) between the point of interest and the UV source. This can be provided directly by the user, or alternatively, can be approximated using the standard relationship between hydrogen column density and visual extinction \cite[e.g][]{guver_ozel_2009}:
\begin{equation}
\label{eq:nH}
    N_\text{H} \approx 2.21 \times 10^{21} A_\text{V}.
\end{equation}
The column densities of CO, N$_2$, and C are then inferred from the total hydrogen column using the local fractional abundance:
\begin{equation}
\label{eq:nX}
    N_\text{X} = N_\text{H} \times \frac{X}{n_\text{gas}},
\end{equation}
where $X$ denotes local number density of the species of interest. The validity of this assumption is discussed in Section \ref{sec:benchmarking}.

Grain-surface chemistry incorporates H$_2$ formation, hydrogenation, freeze-out, thermal desorption, and photodesorption processes. The H$_2$ formation rate on grains follows a temperature-dependent efficiency formalism that accounts for both physisorption and chemisorption binding sites \citep{cazaux_tielens_2002, cazaux_tielens_2004, bosman_2022a}. Hydrogenation, thermal desorption, and photodesorption are implemented following \citet{visser_2011}.

The model also includes cosmic-ray and X-ray ionization processes, with both direct ionization and secondary electron effects \citep{gredel_1987, maloney_1996, yan_1997, stauber_2005}. Simple PAH chemistry is incorporated through charge exchange and electron attachment/detachment processes \citep{lepage_2001, wolfire_2003, jonkheid_2006}, with PAH abundances scaled relative to ISM values. Reactions with vibrationally excited H$_2$ are included following \citet{london_1978, tielens_hollenbach_1985, visser_2018}.

\begin{figure*}
\centering
\includegraphics[clip=,width=1.0\linewidth]{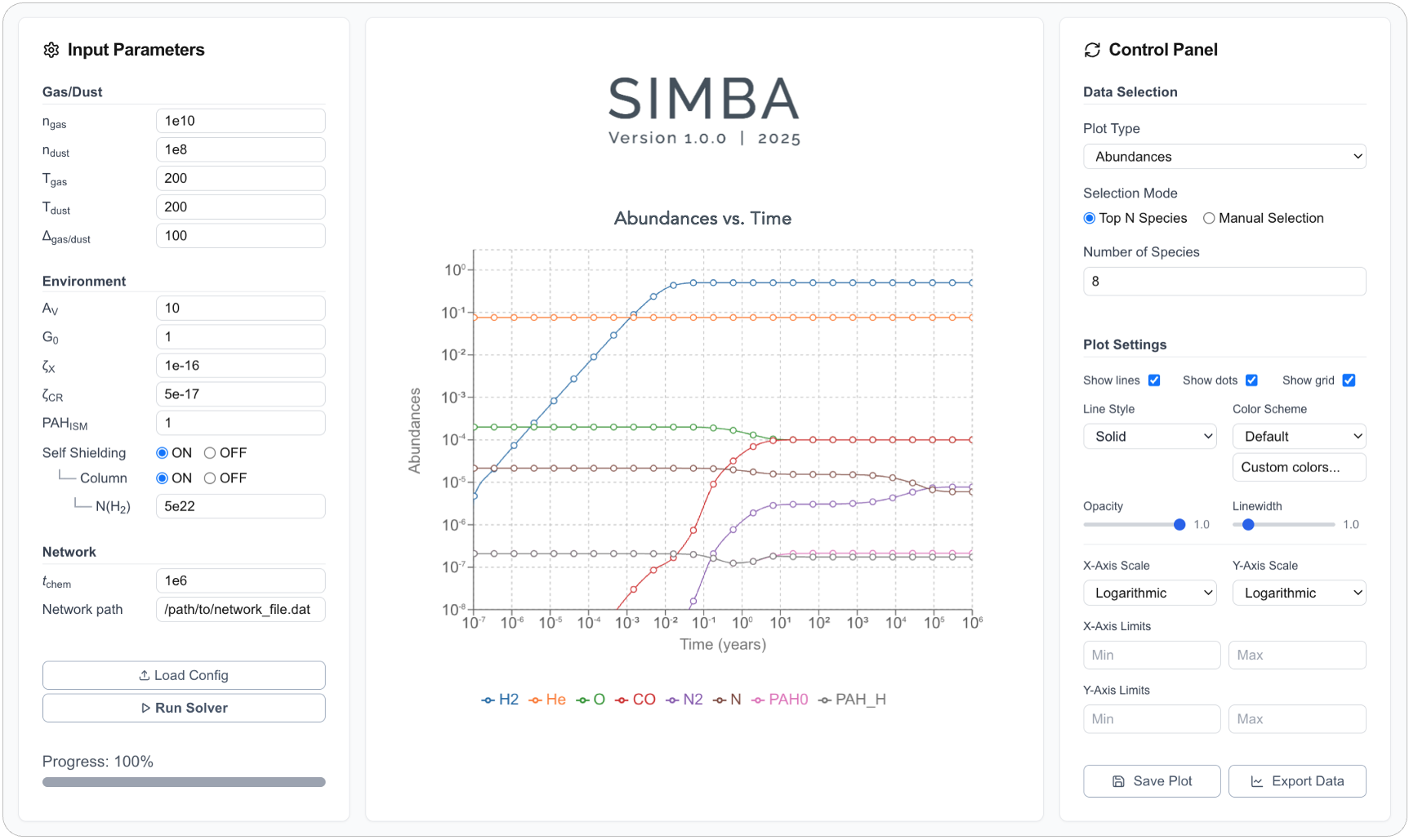}
\caption{Screenshot of the \textsc{simba} graphical user interface (GUI). The GUI consists of the input panel (left), visualisation panel (center), and control panel (right). The GUI\textit{} enables simple analysis of chemical abundances and reaction rates, and provides a wide range of options for plot customisation. Publication-quality images and tabulated data can be exported.}
\label{fig_gui}
\end{figure*}

\subsection{Code implementation}
\label{subsec:code_implementation}

\subsubsection{Overview}
\label{subsubsec:code_overview}

\textsc{simba} is implemented entirely in Python, prioritizing code clarity while achieving acceptable performance through targeted optimisation. The code is structured into several modules that handle distinct aspects of the chemical modelling. This modular design improves maintainability by separating distinct functionalities, and allows users to extend individual components without affecting the rest of the codebase. 

The core data structures are implemented using Python's dataclass framework, which provides a concise way to create classes focused on data storage with built-in validation. The primary dataclasses are:
\begin{itemize}
    \item \texttt{Species}: Contains arrays of abundances, masses, and charges for chemical species, alongside their names
    \item \texttt{Reactions}: Stores reaction networks including reactants, products, rate coefficient parameters, and temperature limits
    \item \texttt{Gas} and \texttt{Dust}: Hold physical properties such as temperatures and densities for each phase
    \item \texttt{Environment}: Contains parameters describing the local conditions like radiation field strength, extinction, and ionization rates
    \item \texttt{Parameters}: Stores simulation settings and physical constants
\end{itemize}
Each dataclass includes validation methods that ensure physical constraints are met (e.g., positive temperatures, valid abundance ranges) and type checking.

Rate coefficients are calculated using a type-based classification system that is deliberately compatible with the established \textsc{dali} code \citep{bruderer_2012, bruderer_2013}. By adopting identical reaction type identifiers, \textsc{simba} can seamlessly utilise chemical networks developed for \textsc{dali} without requiring any reformatting or conversion. This compatibility greatly facilitates network validation and enables direct comparison of results between the two codes, as demonstrated through the detailed benchmarking presented in Section 3. While the code currently reads networks in the \textsc{dali} format, the modular architecture allows straightforward implementation of additional parsers for other network formats, provided the reactions are properly mapped to the appropriate type identifiers. An example network, originally presented in \citealt{keyte_2023}, is distributed with the code.

The core solver is implemented in the \texttt{Simba} class, which manages the chemical network integration. This class employs the \texttt{scipy.integrate.solve\_ivp} routine with the backward differentiation formula (BDF) as the default method, with automatic fallback to Radau if necessary. We adopt relative and absolute tolerances of \texttt{rtol} = $10^{-3}$ and \texttt{atol} = $10^{-8}$ by default, with a further relaxation to \texttt{rtol} = $10^{-2}$ only if both BDF and Radau fail to converge. The maximum step size is restricted to one-tenth of the total integration interval, with the first step set to one ten-thousandth of the starting time. These settings represent a careful balance between numerical stability and computational efficiency, having been validated across a wide range of physical conditions (Section \ref{sec:benchmarking}).

To reduce computational cost, the routines responsible for evaluating the right-hand side of the ODE system and assembling the Jacobian matrix are compiled using \texttt{Numba} in `nopython' mode. \texttt{Numba} is a just-in-time compiler for Python that translates selected functions into optimised machine code, with the `nopython' mode ensuring that all operations are converted to efficient low-level instructions rather than falling back to slower Python object manipulation. The right-hand side calculation iterates through the reaction list, computing the instantaneous formation and destruction terms for each species. Reactions may involve up to three reactants and five products, providing sufficient flexibility for astrochemical applications while maintaining computational tractability.

The Jacobian matrix is assembled explicitly within a \texttt{Numba} kernel as a dense $N \times N$ matrix, where $N$ is the number of species. Each matrix element $\partial f_i/\partial y_j$ represents the sensitivity of species $i$'s formation/destruction rate ($f_i$) to changes in the abundance of species $j$ ($y_j$). Supplying an analytic Jacobian improves both stability and efficiency compared with the finite-difference Jacobians generated internally by the solver, particularly for large chemical networks \citep[e.g.][]{ziegler_2016}. This choice also allows the algebra of the Jacobian to be expressed directly in code, making the code more intuitive and transparent.

The dense Jacobian is subsequently converted to SciPy's LIL (List of Lists) sparse format using \texttt{scipy.sparse.lil\_matrix}. This conversion step incurs computational overhead, but the benefits of sparse representation become increasingly significant as network size grows. Chemical reaction networks exhibit inherent sparsity because each reaction typically involves only a small subset of the hundreds of species present, resulting in Jacobian matrices with sparsity typically exceeding 90\% \citep[e.g.][]{grassi_2013, du_2021, motoyama_2024}. Since sparse solvers operate only on non-zero entries, computational cost scales roughly with the number of non-zero Jacobian elements rather than the full $N^2$ scaling of the dense Jacobian. For typical astrochemical networks, this approach can deliver order-of-magnitude performance improvements 

Specialised physical processes are incorporated through dedicated modules that integrate with the core solver. Self-shielding, for example, is handled by routines that either interpolate tabulated shielding functions (for CO and N$_2$) or apply analytical approximations (for H$_2$ and C). This modular approach allows for straightforward extension to additional species or alternative shielding prescriptions as needed.

Beyond the integration itself, \textsc{simba} includes a comprehensive suite of analysis and visualisation tools. A dedicated analysis module and graphical user interface (Section~\ref{subsubsec:gui}) provide plotting capabilities for species abundances, reaction rates, and chemical pathways, along with utilities for exploring network topology. Results can also be exported in standard formats to facilitate further analysis with external tools.

The overall performance of \textsc{simba} reflects the deliberate design compromise between accessibility and computational efficiency. The primary computational bottlenecks arise from Jacobian conversion from dense to sparse format and the repeated evaluation of rate coefficients at each solver step, which is necessary to capture state-dependent effects such as self-shielding. Despite these limitations, the achieved performance remains well-suited for typical research applications, with integration times of approximately 10 seconds on standard desktop hardware for networks containing 100-200 species and 1000-2000 reactions. This demonstrates that an accessible, Python-based implementation can remain computationally viable for realistic astrochemical modelling applications.

\subsubsection{Installation and usage}

The \textsc{simba} package is available via the Python Package Index (PyPI) and can be installed using the standard \texttt{pip} command:

\begin{verbatim}
pip install simba_chem
\end{verbatim}

\noindent After installation, users can run a chemical simulation by following these steps:

\begin{enumerate}
    \item Import the model and prepare the input files
    \item Initialise the network using the full path to the input file
    \item Run the solver
\end{enumerate}

\noindent An example usage is shown below:

\begin{verbatim}
import simba_chem as simba
network = simba.Simba()
network.init_simba("path/to/input_file.dat")
result = network.solve_network()
\end{verbatim}

\noindent If input or network files are not already available, \textsc{simba} includes helper functions to generate template files that can be modified as needed:

\begin{verbatim}
simba.create_input("path/to/save/input_file.dat")
simba.create_network("path/to/save/network_file")
\end{verbatim}

\noindent Model outputs can be analysed using built-in plotting routines from the \texttt{Analysis} module, or alternatively with custom scripts. For example, to visualise the abundance evolution of selected species:

\begin{verbatim}
analysis = simba.Analysis(network)
analysis.plot_abundance([list_of_species])
\end{verbatim}

\noindent Additional interactive analysis tools are available through the graphical user interface (Section \ref{subsubsec:gui}). Comprehensive documentation, including details on input file structure, configuration parameters, and step-by-step tutorials for running models, is available at:

\begin{verbatim}
simba-chem.readthedocs.io
\end{verbatim}

\begin{figure*}
\centering
\includegraphics[clip=,width=1.0\linewidth]{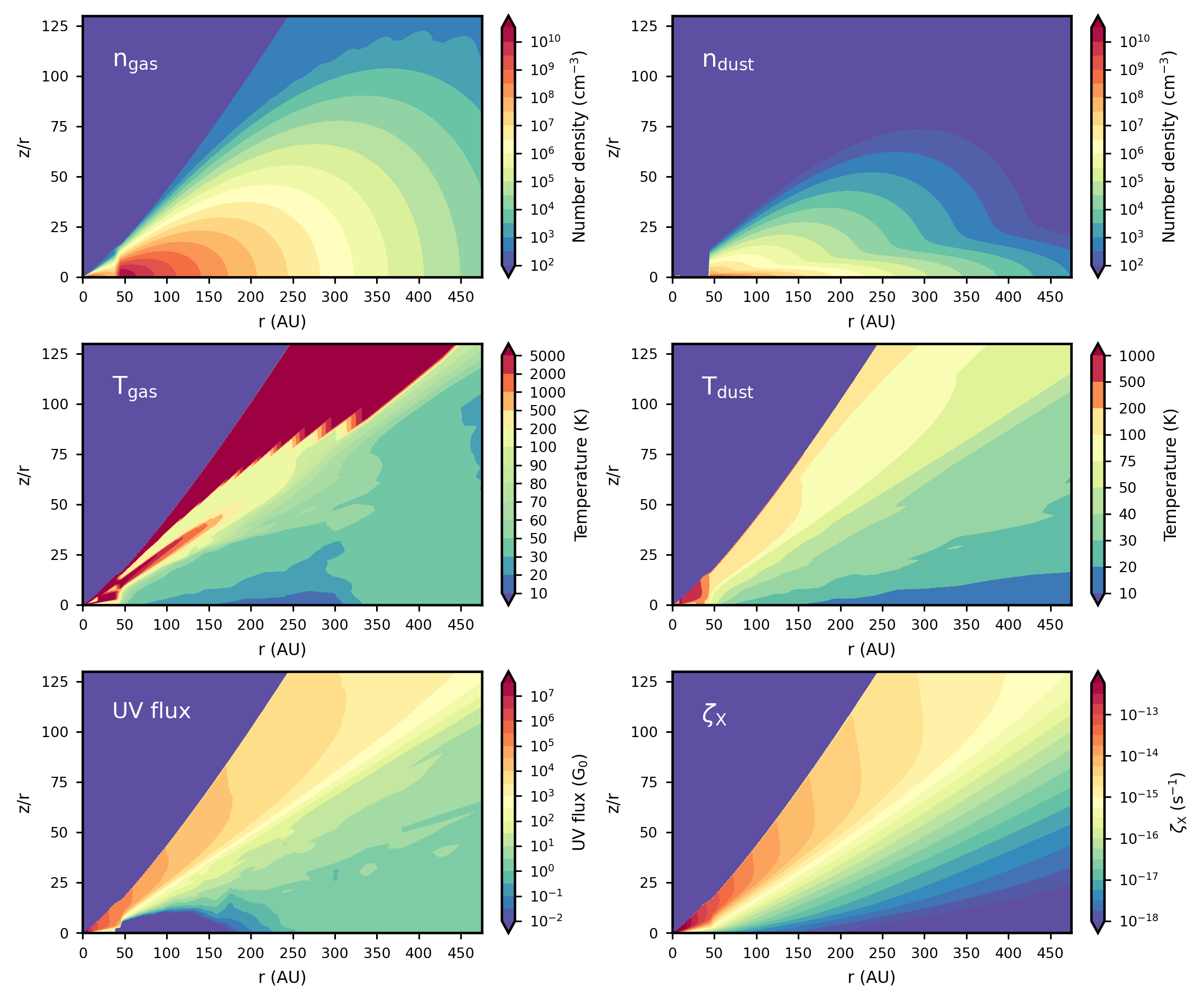}
\caption{Density, temperature, and radiation field structure of the DALI model used for benchmarking the \textsc{simba} code. The DALI model grid consists of 100 radial cells and 80 vertical cells, providing 8000 independent sets of conditions to which the single-point \textsc{simba} model is evaluated against. The full set of parameters used for the DALI model are presented in Table \ref{table:dali_model_parameters}.}
\label{fig_dali_tempdensrad}
\end{figure*}

\subsubsection{Graphical user interface}
\label{subsubsec:gui}

\textsc{simba} features a modern, browser-based graphical interface built in React (Figure \ref{fig_gui}) to streamline the user experience. This interface combines the core solver with comprehensive visualisation tools, enabling users to explore chemical evolution models through an intuitive, interactive environment.

The GUI comes bundled with the standard \texttt{pip} installation and launches locally via the command line:
\begin{verbatim}
simba-gui
\end{verbatim}
The interface requires \texttt{Node.js} (freely available from \url{nodejs.org}) to be pre-installed, but automatically handles all other dependencies during first use.

The user interface organises input parameters through a structured control panel that groups related settings by category: gas and dust properties, environmental conditions, and network specifications. Users can either input parameters manually or load pre-configured files. The solver provides real-time progress updates, and results become immediately available for interactive analysis.

The visualisation panel offers three distinct views: abundance evolution over time, reaction rate evolution, and network pathway diagrams that illustrate chemical connectivity around selected species. Users can choose to display either the most abundant species and fastest reactions automatically, or manually select specific processes for detailed examination. Extensive customisation options allow users to tailor the display, export publication-quality images (SVG), and extract numerical data in standard formats for further analysis.

A significant advantage of \textsc{simba}'s GUI lies in its ability to rapidly explore the parameter space without the computational burden of full multi-dimensional models. While sophisticated codes like \textsc{dali} provide comprehensive modelling capabilities, analysing the vast output from high-resolution models (often hundreds of gigabytes) can be cumbersome. \textsc{simba} is intended to complement, rather than replace, such models by enabling efficient exploration of chemical evolution under varying conditions at specific points of interest. 

This approach makes \textsc{simba} particularly valuable for investigating reaction mechanisms, understanding parameter sensitivities, and analysing localised chemical processes. The intuitive interface also serves as an effective educational tool, allowing students to visualise how different physical conditions influence chemical evolution in real-time, making it an accessible introduction to astrochemistry modelling.

\section{Benchmarking}
\label{sec:benchmarking}

To validate \textsc{simba}'s performance, we conducted comprehensive benchmarking against \textsc{dali} \citep{bruderer_2012, bruderer_2013}, a well-established 1+1D thermochemical modelling code widely used for simulating the physics and chemistry of protoplanetary disks \citep[e.g.][]{kama_2016b, van_der_marel_2016, Facchini17, fedele_2017, cazzoletti_2018, maps_7_bosman_2021, long_2021, leemker_2022, zagaria_2023, stapper_2024, vlasblom_2024, paneque_carreno_2025, vaikundaraman_2025}.

Although \textsc{simba} and \textsc{dali} share similarities in their design, differences between the two codes are expected since they use distinct numerical methods and physical approximations. \textsc{dali} integrates chemical kinetics using the \textsc{limex} linearly implicit extrapolation scheme \citep{limex_ehrig_1999} on a fixed logarithmic time grid with internal error control, while \textsc{simba} relies on \texttt{scipy.integrate.solve\_ivp} using the BDF method and adaptive step sizes. As a result, the two solvers advance the chemistry on different time grids and enforce accuracy through different criteria, which affects how well rapid changes are captured and how numerical errors accumulate. Further differences arise from how the Jacobian is constructed. \textsc{dali} approximates this matrix using finite-difference methods, essentially estimating derivatives by making small numerical perturbations to each variable and observing the resulting changes, while \textsc{simba} calculates the exact analytical derivatives and implements them through \texttt{Numba}-accelerated routines. This difference becomes particularly important when dealing with extremely stiff systems, where reaction timescales span many orders of magnitude, with the analytical approach minimising accumulation of numerical errors that can affect solver stability. Consequently, even when both codes begin with identical physical conditions and chemical abundances, these numerical differences can cause their solutions to gradually diverge as they track chemical evolution.

Beyond the numerics, there are also minor differences in the chemistry implementation. In particular, \textsc{dali} makes direct direct use of the column densities available in its 2D grid to compute self–shielding, while \textsc{simba}, as a single–point model, must approximate these from local conditions or user–supplied values. This approximation can introduce small deviations in the abundances of shielded species such as CO, N$_2$, and C, which may in turn propagate into the broader chemical network. Therefore, while the two codes are broadly consistent in their treatment of the chemistry, small discrepancies are expected, especially for species whose abundances are sensitive to shielding effects.

For this comparison, we configured \textsc{dali} to represent a typical transition disk around a Herbig star, using the parameters detailed in Table \ref{table:dali_model_parameters}. The model employs a chemical network based on a subset of the UMIST 2006 database \citep{woodall_2007}, originally implemented by \citet{bruderer_2012}, which includes 109 species and 1463 individual reactions. The resulting model structure consists of 100 radial cells and 80 vertical cells, creating a total grid of 8000 computational cells. Figure \ref{fig_dali_tempdensrad} shows the spatial distribution of gas and dust densities, temperatures, UV field strength, and X-ray ionization rates throughout this model.

Since \textsc{dali} calculates chemistry across its entire 2D grid while \textsc{simba} operates as a single-point model, we treated each \textsc{dali} grid cell as an independent test case. We extracted the physical and chemical conditions from each of the 8000 \textsc{dali} cells and used these as input parameters for corresponding \textsc{simba} models. The benchmarking assessment was then performed by comparing the predicted chemical abundances between the two codes across all 8000 individual model pairs.

\subsection{Benchmarking metric}

The accuracy of each \textsc{simba} simulation is evaluated through a distance metric, $\delta_\mathrm{cell}$, which is an abundance-weighted species-averaged absolute difference between the \textsc{dali} and \textsc{simba} log abundances:
\begin{equation}
    \delta_\mathrm{cell} = \sum_{s} w(s) \lvert\log_{10}(\frac{X_{\mathrm{DALI}}(s)}{X_{\mathrm{SIMBA}}(s)})\rvert,
\end{equation}
where $X_{\mathrm{DALI}}(s)$ and $X_{\mathrm{SIMBA}}(s)$ are the abundances of species $s$ (relative to hydrogen) predicted by the \textsc{dali} and \textsc{simba} models, respectively. Each data point used in calculating $\delta_\mathrm{cell}$ has a weighting, $w(s)$, applied:
\begin{equation}
    w(s) = \frac{\log_{10}(\frac{X_{\mathrm{DALI}}(s)}{\min(X_{\mathrm{DALI}})}) + 1}{\sum\limits_{s} (\log_{10}(\frac{X_{\mathrm{DALI}}(s)}{\min(X_{\mathrm{DALI}})}) + 1)},
\end{equation}
where $\min(X_{\mathrm{DALI}})$ is the lowest measured abundance in the \textsc{dali} model.
We note that the denominator of each weighting is a normalising term such that $\sum_{s}w(s) = 1$ for all cells, allowing comparison between cells of different conditions. The numerator scales linearly with the measured log \textsc{dali} abundance, with the scale being set by the log-abundance of the least abundant species. For instance, the weightings assigned to \textsc{dali} abundances of $10^{-2}$, $10^{-3}$, and $10^{-6}$ are 0.5, 0.4, and 0.1 respectively. A minimum abundance threshold of $\mathrm{X/H}=10^{-14}$ is imposed to filter out trace species that would contribute more numerical noise than meaningful information to the model comparison.

As an estimate for the difference between \textsc{dali} and \textsc{simba} models, $\delta_\mathrm{cell}$ can be approximated as an uncertainty on the \textsc{simba} abundance, where the $X_\mathrm{DALI}$ can be found within $10^{\pm\delta_\mathrm{cell}} X_{\mathrm{SIMBA}}$. For example, a distance metric $\delta_\mathrm{cell} = 10^{-2}$ is equivalent to a difference of $X_\mathrm{DALI} = {X_\mathrm{SIMBA}}^{+2.3\%}_{-2.3\%}$, averaged across all species.

\subsection{Benchmarking results}
\label{subsec:benchmarking_results}

\begin{figure}
\centering
\includegraphics[clip=,width=1.0\linewidth]{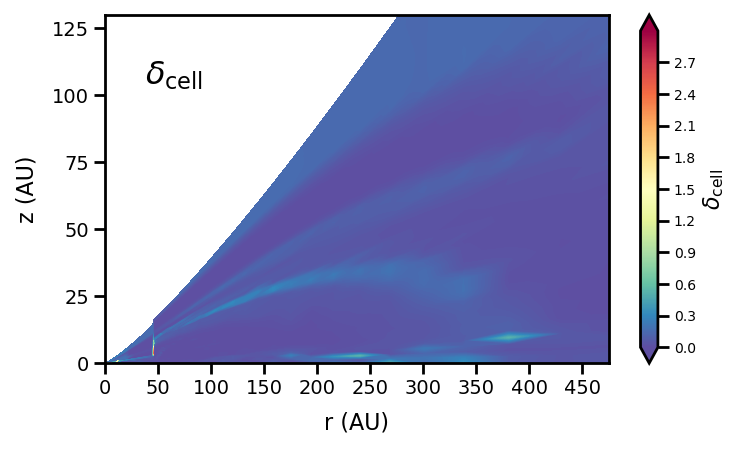}
\caption{Benchmarking comparison between \textsc{simba} and \textsc{dali} chemical abundances across a 2D protoplanetary disk model. The color map displays the distance metric $\delta_\mathrm{cell}$ at each grid position, representing the abundance-weighted absolute difference between \textsc{dali} and \textsc{simba} log abundances. The disk model consists of 100 radial and 80 vertical cells, totalling 8000 individual test cases. Lower $\delta_\mathrm{cell}$ values (darker regions) indicate better agreement between the two codes.}
\label{fig_benchmark_metric}
\end{figure}

The benchmarking results demonstrate excellent overall agreement between \textsc{simba} and \textsc{dali} across the vast majority of test cases, as illustrated by the distance metric distribution shown in Figure \ref{fig_benchmark_metric}. With 7674 successful model completions out of 8000 total cases (95.9\% success rate), \textsc{simba} demonstrates reliable performance across a diverse range of astrophysical conditions. The 326 cases where integration failed to complete are typically located in extremely low-density regions at the upper boundaries of the \textsc{dali} model grid, where physical conditions become unrepresentative of realistic astrophysical environments. 

The distance metric values span from $3.7 \times 10^{-5}$ to $2.4$, with a median of $4.2 \times 10^{-2}$. This median value indicates that typical abundance differences between the two codes are small, corresponding to \textsc{dali} and \textsc{simba} abundances agreeing to within a factor of $\sim 1.1$ when averaged across all species. The majority of cells exhibit low $\delta_\mathrm{cell}$ values (darker regions in Figure \ref{fig_benchmark_metric}), confirming that significant discrepancies are highly localised rather than representing systematic limitations of the \textsc{simba} implementation.

In regions where the distance metric is poorest, discrepancies primarily stem from \textsc{simba}'s approximation of molecular column densities used in self-shielding calculations (Equations \ref{eq:nH} and \ref{eq:nX}). While \textsc{simba} allows users to specify the total hydrogen column density as an input parameter, the column densities for other molecules involved in self-shielding (N$_2$, C, CO) must be inferred from their local abundances. In contrast, \textsc{dali} can calculate the exact column density for all species to any cell by integrating along sight lines through its full 2D grid. This fundamental limitation of single-point models means that \textsc{simba}'s accuracy is affected most significantly in regions where molecular abundances change rapidly over small spatial scales, as the local approximation becomes increasingly inadequate. This effect is most pronounced at sharp transitions, such as the edge of the dust and gas cavity at $r=4$5 au, where density and radiation fields undergo dramatic variations that cannot be properly represented by local conditions alone. Nevertheless, these localised difference represent less than 5\% of the total parameter space, confirming that \textsc{simba}'s single-point approximation remains effective for most astrophysical applications. Comparisons between the chemical evolution of \textsc{simba} and \textsc{dali} are visualised in Figure \ref{fig_benchmark_comparison} at three representative locations. The selected cells are located at $(60.0, 4.9)\; \text{au}$, $(120.8, 35.6)\; \text{au}$ and $(0.44, 0.02)\; \text{au}$, corresponding to $10^{\text{th}}$, $50^{\text{th}}$, and $100^{\text{th}}$ percentile distance metric values respectively. Figure \ref{fig_metric_distribution} presents the distribution of metric values across all instances of successful integration. This distribution is heavily skewed to lower values, emphasising that the majority of the parameter space is well simulated by \textsc{simba}.

\section{Chemical evolution of a photoevaporative flow}
\label{sec:example}

To demonstrate \textsc{simba}'s capabilities, we present a model that simulates chemical evolution in a photoevaporative flow. This flow originates from the outer edge of a protoplanetary disk and represents the conditions found in externally irradiated disks within star-forming regions like the Orion Nebular Cluster. In these environments, intense far-ultraviolet (FUV) radiation from nearby massive stars drives the photoevaporative flows \citep[e.g.][]{odell_wen_1994, bally_2000}. While FUV-driven photoevaporation represents only one mechanism of disk dispersal, with magneto-thermal winds from non-ideal MHD processes also playing significant roles in shaping accretion and outflow geometries \citep[e.g.][]{bai_2017, gressel_2020}, the photoevaporative scenario provides an ideal testbed for exploring chemistry with \textsc{simba}.

\subsection{Model setup}

\begin{figure*}
\centering
\includegraphics[clip=,width=1.0\linewidth]{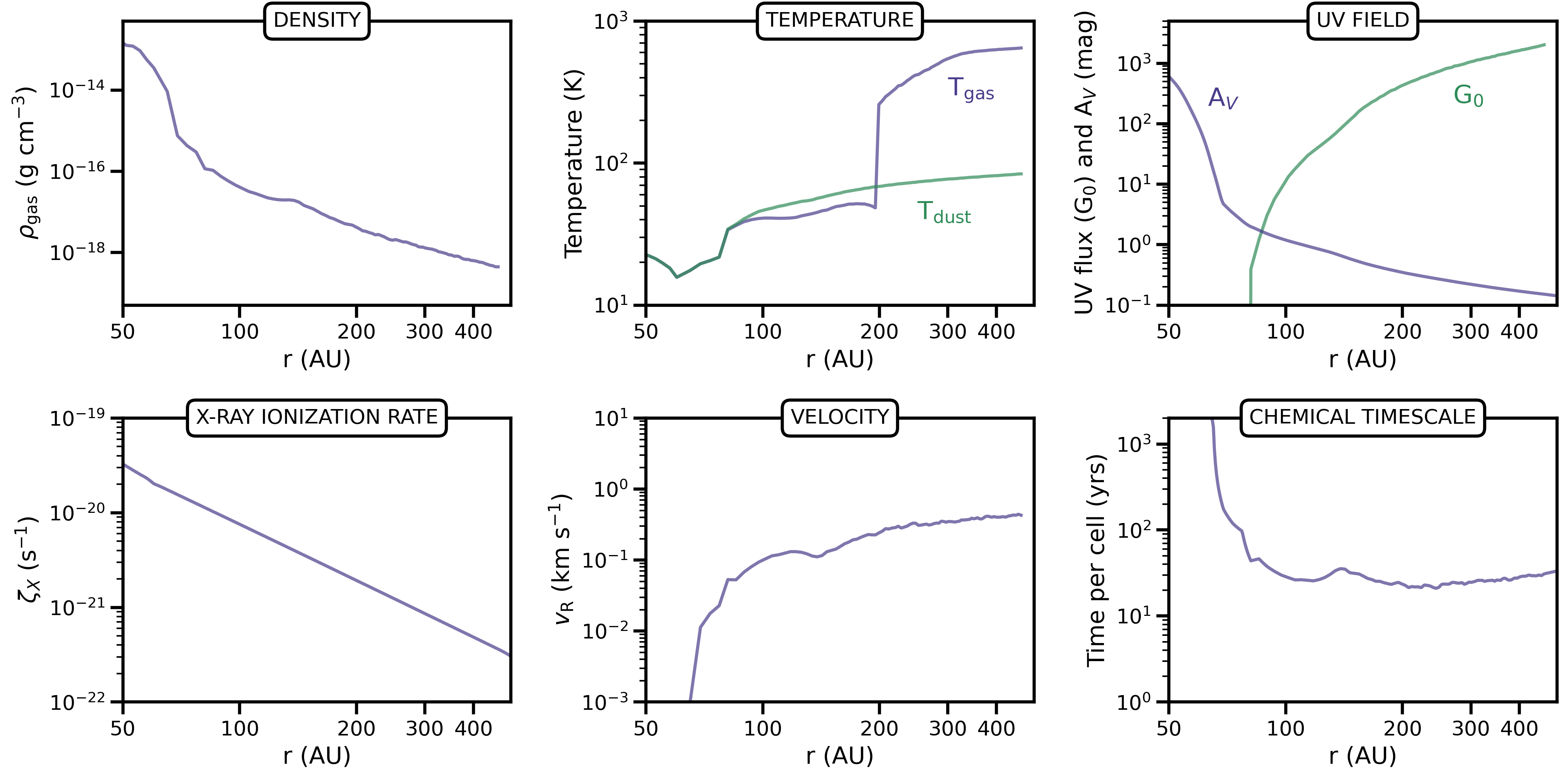}
\caption{Physical parameter profiles for the 1D photoevaporative flow model. The six panels show the radial variation of key parameters from the disk outer edge (50~au) to the domain boundary (500~au). \emph{Top row:} Gas density ($\rho_{\rm gas}$), gas and dust temperatures ($T_{\rm gas}$, $T_{\rm dust}$), and UV field conditions (A$_V$, G$_0$). \emph{Bottom row:} X-ray ionization rate ($\zeta_X$), radial gas velocity ($v_R$), and chemical evolution timescale per grid cell. The structure represents a typical externally irradiated disk with photoevaporative mass loss driven by intense FUV radiation (5000~$G_0$), adopted from \citet{ballabio_2023} and \citet{keyte_haworth_2025}.}
\label{fig_modelparams}
\end{figure*}

To extend the standard single-point \textsc{simba} model to higher dimensions, we establish a model grid where each cell has unique conditions and runs a separate simulation. In this study, we focus on the midplane of a protoplanetary disk using a 1D radial grid. We populate this grid with physical parameters from a representative chemical model \citep{keyte_haworth_2025}, which characterises a typical externally irradiated disk (model parameters are listed in Table \ref{table:fried_model_parameters}). The density structure is adopted from \citet{ballabio_2023}, who generated these profiles using the photochemical hydrodynamics code \textsc{torus-3d-pdr} \citep{bisbas_2015}. Our computational domain extends from 50 au (the disk outer edge where the photoevaporative wind launches) to 500 au at the domain boundary. The radial variation profiles for all physical parameters are shown in Figure \ref{fig_modelparams}.

Having established the physical parameter grid, we implement individual \textsc{simba} models within each grid cell to determine chemical abundances at each location. Our simulations use the chemical network from \citet{bruderer_2012} and adopt initial abundances based on \citet{ballering_2021}. These abundances represent typical molecular cloud compositions with a Solar-like C/O ratio and depleted levels of carbon and oxygen, similar to those inferred in several observed disks \citep[e.g.][]{maps_7_bosman_2021}. While we focus on Solar-like conditions here, we note that metallicity can significantly influence photoevaporation efficiency. Lower-metallicity disks disperse more rapidly due to reduced dust shielding and altered heating-cooling balance \citep{nakatani_2018}. Exploring how such metallicity variations affect the coupled chemical and dynamical evolution of photoevaporative flows is an important avenue for future work. In this study, however, we adopt a fixed set of abundances as a representative baseline for Solar-like disks. The complete set of initial abundances is provided in Table \ref{table:initial_abundances}.

We employ two distinct methodological approaches. In our first approach, we execute a 'static' simulation where each cell initialises with identical elemental and molecular abundances. Chemistry then evolves uniformly across all cells for a consistent timescale of 1 Myr, corresponding to the typical age of such systems. As a complementary approach, we implement a simplified treatment of gas dynamics by utilizing the sequential nature of the 1D \textsc{simba} framework. In this `dynamic' simulation, we initially evolve chemical abundances in the first cell at the wind base following the static approach, evolving for 1 million years to represent typical disk abundances at this location. Subsequently, these final abundances propagate to adjacent cells as initial conditions, evolving further according to the gas crossing time between cell boundaries, calculated using the gas velocity given by:
\begin{equation}
    v_\text{R} = \frac{\dot{M}}{4 \pi R^2 \mathcal{F} \rho},
\end{equation}
where $\dot{M}$ is the mass accretion rate, $R$ is the distance form the central star, $\mathcal{F}$ is the fraction of solid angle subtended by the disk outer edge \citep{adams_2004}, and $\rho$ is the gas density.

Each successive cell inherits the composition from its predecessor, with chemistry further evolving for a duration equal to the local crossing time, continuing until reaching the computational domain boundary. The gas velocity and chemical evolution timescales as a function of radius are illustrated in Figure \ref{fig_modelparams}. While this `dynamic' approach adopts a highly simplified treatment of gas dynamics, it provides a valuable first-order approximation enabling insightful comparison with the standard 'static' model. A similar approach as been used by e.g. \citet{cridland_2025} to study the chemical evolution of material accreting on to a circumplanetary disk. Analyses of both approaches are presented in subsequent subsections.

\begin{table}
\caption{Parameters adopted for our showcase 1D model.}             
\label{table:fried_model_parameters}      
\centering
\begin{tabular}{l l l}     %
\hline\hline       
Parameter & Description & Value \\ 
\hline                    
   $M_*$                & Stellar mass                    & $1.0$ M$_\odot$     \\
   F$_\text{FUV}$       & Integrated FUV field strength   & $5000$ G$_0$        \\
   $\Sigma_\text{1au}$  & Surface density at $r=1$ au     & 1000 g cm$^{-2}$    \\
   $R_\text{d}$         & Disk outer edge                 & 50 au               \\
   $\dot{M}$            & Mass accretion rate             & $4.35 \times 10^{-8}$ M$_\odot$ yr$^{-1}$   \\
   $\mathcal{F}$        & Angle subtended by disk edge    & 0.12 rad     \\
\hline                  
\end{tabular}
\end{table}

\subsection{Results}
\label{sec:results}

\subsubsection{Survival of molecular ices}

\begin{figure*}
\centering
\includegraphics[clip=,width=0.88\linewidth]{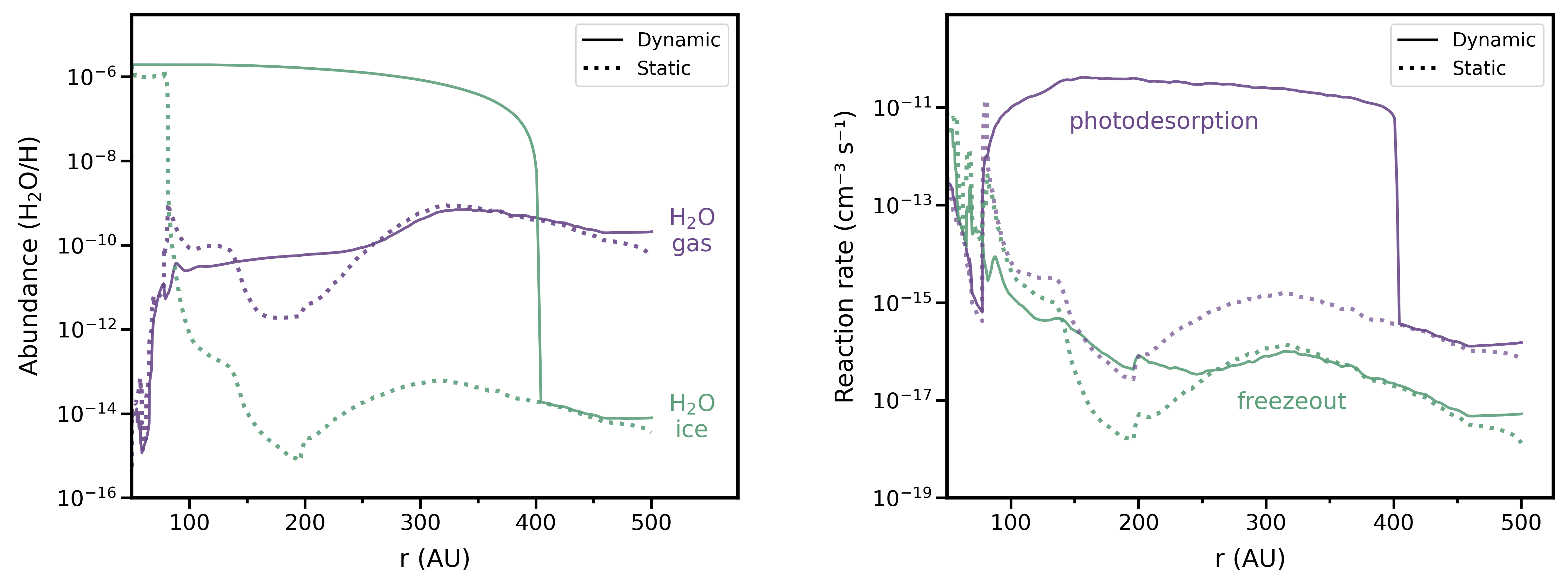}
\caption{\emph{Left:} Radial distribution of H$_2$O abundances in both ice and gas phases derived from our \textsc{simba} simulations. Solid lines denote results from the `dynamic' model, while dotted lines represent the `static' model. The dynamic model demonstrates significantly extended survival of H$_2$O ice at higher abundances throughout the photoevaporative flow, whereas gas-phase H$_2$O abundances remain similar between both modelling approaches. \emph{Right:} Corresponding photodesorption and freezeout reaction rates for both models, which are the dominant processes governing the abundance distributions.}
\label{fig_h2o}
\end{figure*}

We first examine the survival of molecular ices within the photoevaporative flow, beginning with H$_2$O as a representative example. Figure \ref{fig_h2o} (left panel) illustrates the radial distribution of H$_2$O abundances in both ice and gas phases, derived from our `static' and `dynamic' models. At the base of the wind ($r=50$ au), H$_2$O ice is the primary oxygen carrier in both simulations, with an abundance of H$_2$O/H = $1.94 \times 10^{-6}$.

The two models exhibit markedly different behaviour moving outward into the flow. In the static model, H$_2$O ice abundance decreases sharply with radius, showing a dramatic drop at approximately $r \sim 80$ au. This decline directly corresponds to the rapid decrease in gas density and extinction at this location, combined with increasing external FUV flux. In contrast, the dynamic model demonstrates significantly extended H$_2$O ice survival, maintaining elevated abundances until $r \sim 400$ au. This enhanced persistence occurs because material flows outward faster than photodesorption can destroy the ices. This phenomenon of advection-dominated survival has also been demonstrated for gas-phase molecules in photoevaporative winds, where H$_2$ can persist far beyond its photodissociation radius due to rapid material transport \citep{sellek_2024}.

The mechanisms controlling H$_2$O ice formation and destruction are illustrated in Figure \ref{fig_h2o} (right panel), which presents the photodesorption and freezeout rates for both models. The pronounced decrease in H$_2$O ice abundance at $r \sim 400$ au in the dynamic model results from the photodesorption rate ($k_\text{pd}$) dependence on total ice abundance:
\begin{equation}
k_\text{pd} = \pi a_\text{gr}^2 n_\text{gr} f(X) Y(X) F_0 \exp(-\tau_\text{UV,eff}),
\end{equation}
where $a_\text{gr}$ denotes grain radius, $n_\text{gr}$ represents grain number density, $Y(X)$ signifies the photodesorption yield, $F_0$ corresponds to the unattenuated FUV flux, and $\tau_\text{UV,eff}$ represents effective UV extinction. The dimensionless coefficient $f(X)$ is defined by:
\begin{equation}
f(X) = \frac{n_\text{s}(X)}{\text{max}(n_\text{ice}, N_\text{b}n_\text{gr})},
\end{equation}
where $n_\text{s}(X)$ indicates the number density of ice species $X$, $n_\text{ice}$ represents the total ice number density, and $N_\text{b}=10^6$ corresponds to the characteristic number of binding sites per grain \citep{visser_2011}. Thus, as the cumulative ice abundance progressively diminishes with increasing distance into the flow, a critical threshold is eventually reached. Below this threshold, the remaining ice abundance becomes sufficiently low that all residual ices undergo rapid, complete desorption.

Despite the substantial differences in H$_2$O ice abundances between the two models, gas-phase H$_2$O abundances remain remarkably similar throughout the flow. This is because, in the static model, desorbed H$_2$O ice undergoes photodissociation on timescales much shorter than the chemical evolution timescale of 1 Myr:
\begin{gather}
\text{H}_2\text{O} + h\nu \rightarrow \text{OH} + \text{H}.
\end{gather}
The OH radicals formed via this pathway are subsequently destroyed through photodissociation and gas-phase reactions with atomic nitrogen:
\begin{gather}
\text{OH} + h\nu \rightarrow \text{O} + \text{H}, \\
\text{OH} + \text{N} \rightarrow \text{NO} + \text{H}.
\end{gather}
This results in atomic oxygen becoming the dominant oxygen carrier in the static model beyond $\sim 80$ au.

\begin{figure}
\centering
\includegraphics[clip=,width=0.95\linewidth]{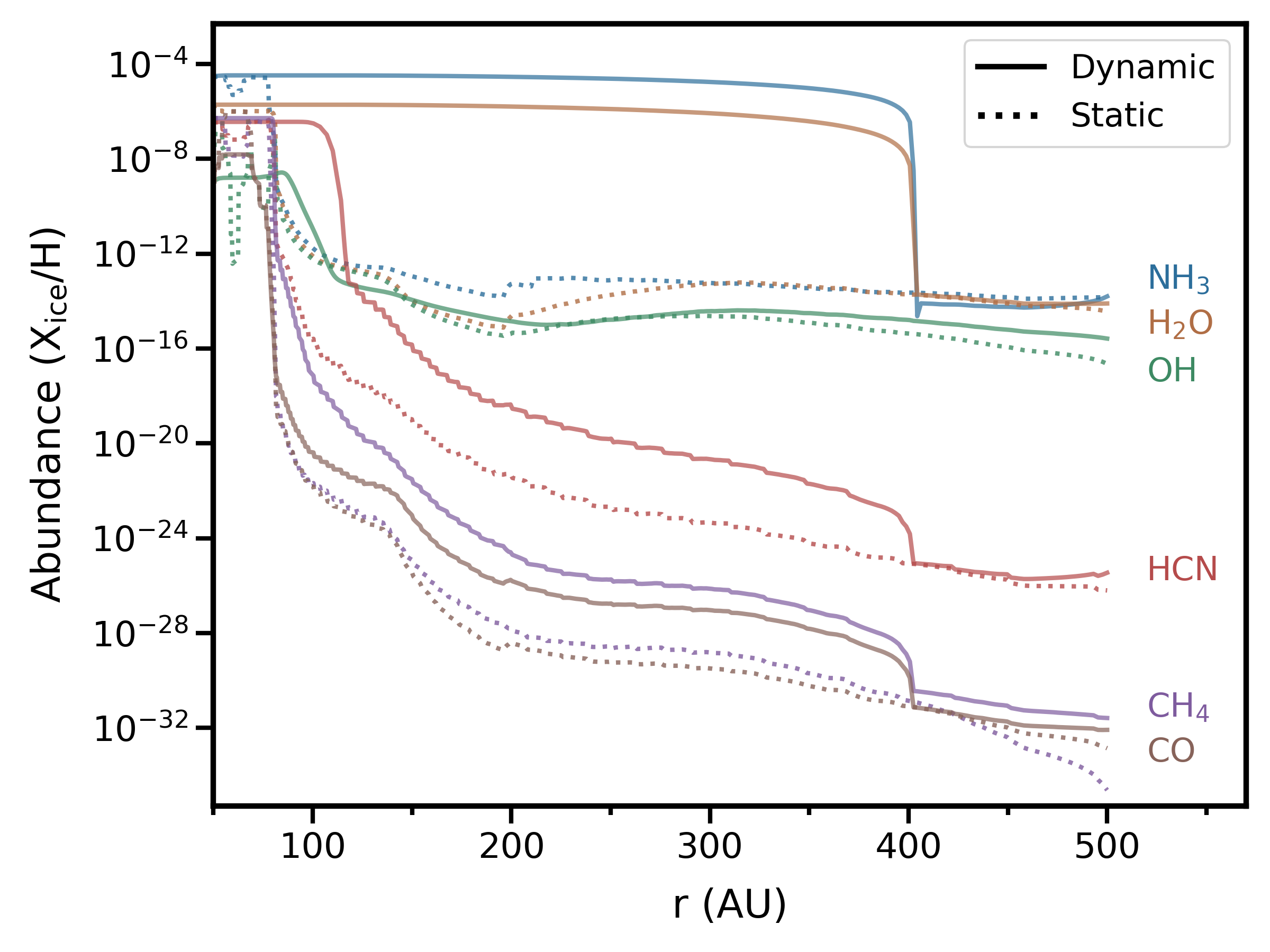}
\caption{Radial abundance distributions of molecular ices from our \textsc{simba} simulations, comparing the `dynamic' model (solid lines) and `static' model (dotted lines). High binding energy species (H$_2$O, NH$_3$) persist far into the photoevaporative flow in the dynamic model due to negligible thermal desorption at low dust temperatures. In contrast, low binding energy species (CO, CH$_4$, HCN) show similar abundance profiles in both models because their depletion is governed primarily by thermal desorption, which operates on shorter timescales than the flow dynamics can counteract.}
\label{fig_ices}
\end{figure}

The enhanced survival of H$_2$O ice in the dynamic model can be partially attributed to its relatively high binding energy, which effectively prevents significant thermal desorption. Dust temperatures throughout the flow remain below 100 K (Figure \ref{fig_modelparams}), substantially lower than water ice's canonical sublimation temperature of $T\text{sub} \approx 150$ K. This temperature dependence explains why molecules with comparably high binding energies, such as NH$_3$, exhibit similar behaviour to H$_2$O ice in both models, with NH$_3$ ice abundance also remaining elevated in the dynamic model out to $r\sim 400$ au (Figure \ref{fig_ices}).

In contrast, the ice-phase abundance of species with significantly lower binding energies than H$_2$O, including CO, CH$_4$, and HCN, are principally governed by thermal desorption rather than photodesorption. Since thermal desorption operates on shorter timescales than the crossing timescales, these species show more comparable ice phase abundances between the static and dynamic models. Their ice abundances diminish at similar radii in both model scenarios, as shown in Figure \ref{fig_ices}.

\subsubsection{Volatile carbon, oxygen, and nitrogen}

Our analysis of molecular ice survival revealed that species with higher binding energies like H$_2$O and NH$_3$ persist much further into the photoevaporative flow in the dynamic model compared to the static model. This extended ice survival fundamentally alters the elemental budget available for gas-phase chemistry, as a significant fraction of heavy elements remains locked in solid form rather than participating in gas-phase reactions. We now examine how this affects the distribution of major volatile carbon, oxygen, and nitrogen carriers throughout the flow.

\begin{figure*}
\centering
\includegraphics[clip=,width=0.95\linewidth]{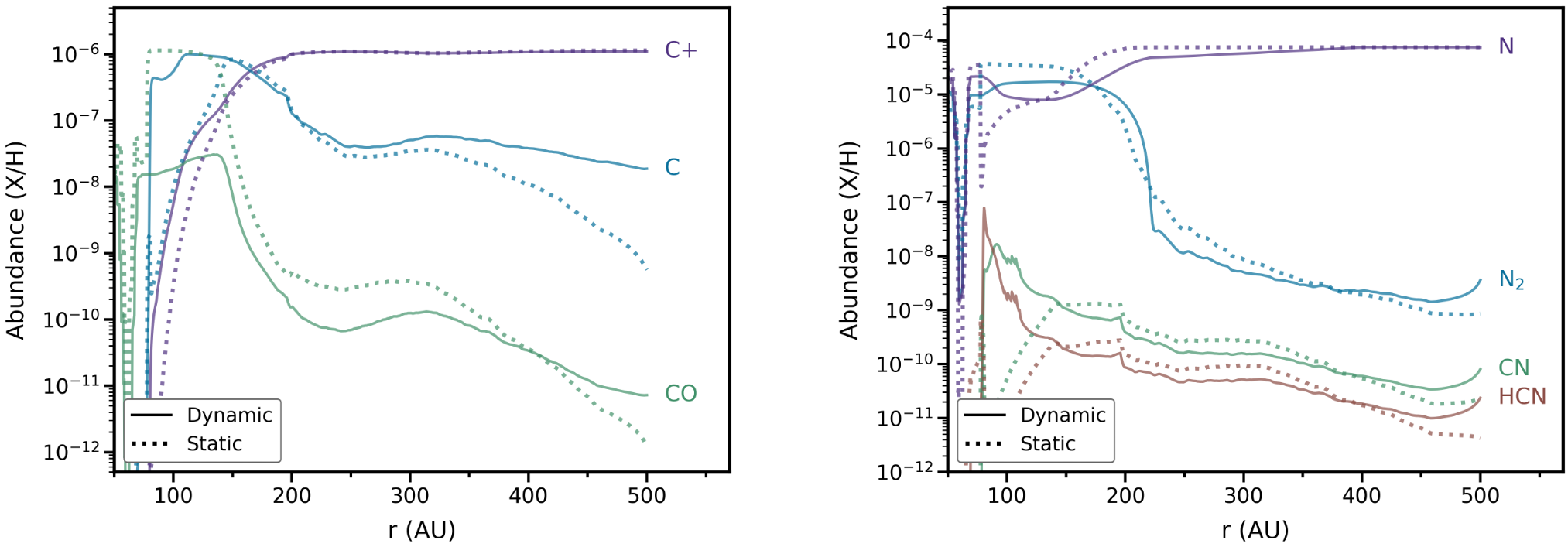}
\caption{Radial abundance distributions of major gas-phase carbon carriers (left panel: CO, C, C$^+$) and nitrogen carriers (right panel: N, N$_2$, CN, HCN) derived from our \textsc{simba} simulations. Solid lines represent the `dynamic' model while dotted lines show the `static' model. The dynamic model shows significantly different carbon and nitrogen chemistry in the inner regions ($r < 170$ au) due to extended NH$_3$ ice survival, with carbon remaining predominantly atomic rather than forming CO. Beyond $\sim 170$ au, both models converge as the intense external UV field dominates the chemical evolution regardless of initial formation pathways.}
\label{fig_carbon_nitrogen}
\end{figure*}

Figure \ref{fig_carbon_nitrogen} presents the radial abundance profiles for some key gas-phase carbon carriers (CO, C, and C$^+$) in the left panel and nitrogen carriers (N, N$_2$, CN, HCN) in the right panel. At the wind base ($r = 50$ au), most volatile carbon is initially locked in CH$_4$ ice and HCN ice, while nitrogen is distributed primarily between NH$_3$ ice and gas-phase N and N$_2$ (Table \ref{table:initial_abundances}). As material flows outward from the disk edge, both models first experience rapid thermal desorption of the most volatile ices. CH$_4$ and HCN ices become thermally desorbed by approximately 70 au in both models, releasing substantial amounts of carbon into the gas phase for subsequent chemical processing. This similarity between models occurs because thermal desorption operates on timescales much shorter than the flow crossing times, making it largely independent of the dynamic treatment.

The chemical evolution following this initial desorption phase differs dramatically between the two approaches. In the static model, the carbon released from thermally desorbed CH$_4$ and HCN ices follows a predictable sequence of photochemical processing. The desorbed carbon initially converts to CO, which remains stable and survives out into the flow to around 140 au. At this location, rapidly decreasing gas densities and increasing external UV flux create conditions where CO self-shielding becomes ineffective. Beyond 140 au, photodissociation destroys CO, making atomic carbon the dominant carbon carrier until around 170 au. Further outward, continued exposure to the intense external UV field photoionises the atomic carbon, leaving C$^+$ as the primary carbon carrier from this point to the flow boundary. This radial progression, from molecular to atomic to ionised forms, represents the expected behaviour for static models and aligns with predictions from previous photoevaporation studies \citep[e.g.][]{haworth_clarke_2019}.

The dynamic model exhibits markedly different behaviour due to the extended survival of NH$_3$ ice. While the initial thermal desorption of CH$_4$ and HCN ices proceeds similarly to the static case, the subsequent chemical evolution diverges significantly. The persistence of NH$_3$ ice far into the flow results in low gas-phase N$_2$ abundances throughout most of the domain. Under these nitrogen-limited conditions, the carbon released from CH$_4$ and HCN ices participates in different reaction pathways that consume nitrogen to form N$_2$, rather than being converted to CO.

This chemistry proceeds through a series of photochemical reactions that couple carbon and nitrogen processing. Initially, photodissociation breaks down CH$_4$:
\begin{gather}
\text{CH}_4 + h\nu \rightarrow \text{CH}_2 + \text{H}_2.
\end{gather}
The resulting CH$_2$ radical reacts with available atomic nitrogen:
\begin{gather}
\text{CH}_2 + \text{N} \rightarrow \text{HCN} + \text{H}.
\end{gather}
This reaction, combined with HCN already desorbed from ices, provides a substantial HCN reservoir. Subsequent photodissociation of HCN:
\begin{gather}
\text{HCN} + h\nu \rightarrow \text{CN} + \text{H},
\end{gather}
produces CN radicals that can be destroyed through multiple pathways:
\begin{gather}
\text{CN} + h\nu \rightarrow \text{N} + \text{C},\\
\text{CN} + \text{N} \rightarrow \text{N}_2 + \text{C}.
\end{gather}
These reaction sequences effectively convert the carbon budget from desorbed ices into N$_2$ and atomic carbon, rather than the CO formation seen in the static model. This fundamental difference demonstrates how the survival of ices can reshape the entire chemical evolution of the photoevaporative flow.

The elemental oxygen budget follows a similar pattern. At the wind base, almost all oxygen is initially locked into H$_2$O ice. In the static model, most of this oxygen is desorbed by around 70 au, making it available for gas-phase chemistry. CO becomes the major oxygen carrier until around 140 au, where photodissociation converts it to atomic oxygen, which remains the main gas-phase oxygen carrier to the domain edge. In contrast, the dynamic model retains most oxygen as H$_2$O ice far into the flow, leaving relatively little oxygen available for gas-phase chemistry. This contributes to the lower CO abundances discussed previously. However, similar to the carbon case, the oxygen carriers in both models converge by around 170 au, where photodissociation of CO drives the atomic oxygen budget in both scenarios.

Despite these substantial differences in the inner regions, both models converge to similar abundances beyond approximately 170 au. At these large distances, the intense external UV field becomes the dominant driver of chemical evolution, overwhelming the initial differences in elemental budgets between the two approaches. The carbon species in both models follow similar photochemical pathways, with atomic carbon becoming photoionised to form C$^+$ as the primary carbon carrier. Similarly, nitrogen species abundances become comparable between the models as the harsh radiation environment controls the chemical balance regardless of the initial formation history.

The contrast between these scenarios highlights the importance of considering realistic timescales and material transport when modelling chemical evolution in dynamic astrophysical environments. The static assumption of independent chemical evolution in each spatial location may significantly underestimate the survival of molecular ices and consequently misrepresent the gas-phase chemical composition throughout the flow. These findings suggest that traditional static chemical models may need to be reconsidered for environments where material transport timescales are comparable to or shorter than chemical processing timescales.

\section{Summary}

We have presented \textsc{simba}, a new Python-based single-point astrochemical modelling package designed to solve chemical reaction networks across diverse astrophysical environments. The software combines accessibility and ease-of-use with computational efficiency through targeted optimisation, making it particularly valuable for researchers new to astrochemistry and as an educational tool. Key contributions and findings of this work include:
\begin{enumerate}
    \item \textsc{simba} provides a comprehensive Python-based framework for chemical modelling that prioritises code clarity and user accessibility while maintaining suitability for research. The package includes a modern browser-based graphical interface, extensive documentation, and compatibility with established chemical networks, lowering barriers to entry for newcomers to the field.
    \item Comprehensive benchmarking against the established \textsc{DALI} code demonstrates \textsc{simba}'s accuracy across a diverse range of astrophysical conditions, validating its suitability for quantitative research applications.
    \item Our illustrative case of modelling a photoevaporative flow reveals significant differences between traditional static approaches and simplified dynamical treatments. Ice-phase species with relatively high binding energies (H$_2$O, NH$_3$) can survive much further into the flow when material transport is considered, fundamentally altering the gas-phase elemental budget and chemical evolution.
    \item The extended survival of molecular ices in dynamic environments impacts gas-phase chemistry, with the static assumption potentially misrepresenting gas-phase abundances by several orders of magnitude in the inner regions of flows. These findings highlight the importance of considering realistic timescales and material transport in astrochemical models, particularly in dynamic environments where chemistry and physics are coupled.
\end{enumerate}
The \textsc{simba} package is freely available as open-source software, with documentation and examples to support users in the astrochemistry community. By balancing ease of use with computational performance, \textsc{simba} should prove valuable for both research and education.

\section*{Acknowledgements}

We thank the reviewer for their helpful and insightful comments. We thank Simon Bruderer for his support. LK is funded by UKRI guaranteed funding for a Horizon Europe ERC consolidator grant (EP/Y024710/1). JR is supported by the STFC UCL Centre for Doctoral Training in Data Intensive Science (grant ST/W00674X/1) including departmental and industry contributions.

\section*{Data Availability}

The GitHub repository for \textsc{simba} can be found at \url{https://github.com/lukekeyte/SIMBA/}.



\bibliographystyle{mnras}
\bibliography{bibliography} 




\appendix

\newpage

\section{DALI benchmarking model}

\begin{table*}
\caption{DALI disk model parameters used for benchmarking.}             
\label{table:dali_model_parameters}      
\centering
\begin{tabular}{l l l}     %
\hline\hline       
                      
Parameter & Description & Fiducial\\ 
\hline                    
   \rsub                & Sublimation radius                         & 0.2 au    \\
   \rgap                & Inner disk size                            & 10 au     \\
   \rcav                & Dust cavity radius                         & 45 au      \\
   $R_c$                & Critical radius for surface density        & 50 au      \\
   $\delta_\text{gas}$  & Gas depletion factor inside \rcav          & $10^{-3}$  \\
   $\delta_\text{dust}$ & Dust depletion factor inside \rcav         & $10^{-5}$  \\
   $\gamma$             & Power law index of surface density profile & 1.0        \\
   $\chi$               & Dust settling parameter                    & 0.2        \\
   $f$                  & Large-to-small dust mixing parameter       & 0.85       \\
   $\Sigma_c$           & $\Sigma_\text{gas}$ at $R_c$               & 17.0 g cm$^{-2}$  \\
   $h_c$                & Scale height at $R_c$                      & 0.06 rad      \\
   $\psi$               & Power law index of scale height            & 0.2       \\
   \gasdust             & Gas-to-dust mass ratio                     & 100        \\
   $L_*$                & Stellar luminosity                         & $10\; L_\odot$        \\
   $T_*$                & Stellar temperature                        & $10000$ K        \\
   $L_X$                & Stellar X-ray luminosity                   & $1.0 \times 10^{30} \text{ erg s}^{-1}$    \\
   $T_X$                & X-ray plasma temperature                   & $7.0 \times 10^{7}$ K     \\
   $\zeta_\text{cr}$    & Cosmic ray ionization rate                 & $5.0 \times 10^{-17}$ s$^{-1}$  \\
   $M_\text{gas}$       & Disk gas mass                              & $1.21 \times 10^{-2}$ \msun   \\
   $M_\text{dust}$      & Disk dust mass                             & $1.21 \times 10^{-4}$ \msun   \\
   $t_\text{chem}$      & Timescale for time-dependent chemistry     & \text{1 Myr} \\
\hline                  
\end{tabular}
\end{table*}

\section{1D photoevaporative flow model parameters} 

\begin{table*}
\caption{Abundances for the 1D photoevaporative flow model. Both the `static' and `dynamical' models are initialised with the abundances listed in the `Initial' column at $r=50$ au. The static model applies these initial abundances uniformly across all grid cells and uses a fixed chemical evolutionary timescale of 1 Myr throughout. In contrast, the dynamical model uses these abundances only at $50$ au, with each subsequent cell inheriting its initial state from the final state of the preceding cell; its chemical timescale is determined by each cell's size and gas velocity. The `Final' column shows the resulting abundances at $r=50$ au for both models after chemical processing. For clarity, only species with abundances X/H $>10^{-12}$ are shown.}             
\label{table:initial_abundances}      
\centering
\begin{tabular}{l l l @{\hspace{2cm}} l l l}
\hline\hline        
\multicolumn{3}{c|}{\emph{Gas Phase}} & \multicolumn{3}{c}{\emph{Ice Phase}} \\
\hline
Species & Initial [X/H] & Final [X/H] at wind base & Species & Initial [X/H] & Final [X/H] at wind base \\ 
\hline
H$_2$      & $5.00 \times 10^{-1}$ & $5.00 \times 10^{-1}$     & NH$_{3{\rm ,ice}}$   & - & $3.11 \times 10^{-5}$ \\
He         & $9.00 \times 10^{-2}$ & $9.00 \times 10^{-2}$     & H$_2$O$_{\rm ice}$   & - & $1.94 \times 10^{-6}$ \\
N$_2$      & $3.51 \times 10^{-5}$ & $1.22 \times 10^{-5}$     & CH$_{4{\rm ,ice}}$   & - & $5.24 \times 10^{-7}$ \\
NH$_3$     & $4.80 \times 10^{-6}$ & $2.38 \times 10^{-12}$    & HCN$_{\rm ice}$      & - & $3.63 \times 10^{-7}$ \\
CO         & $1.00 \times 10^{-6}$ & $2.07 \times 10^{-8}$     & HNC$_{\rm ice}$      & - & $1.91 \times 10^{-7}$ \\
H$_2$O     & $1.00 \times 10^{-6}$ & - & N$_{\rm ice}$         & - & $1.68 \times 10^{-7}$ \\
PAH        & $6.00 \times 10^{-7}$ & $1.61 \times 10^{-7}$     & N$_{2{\rm ,ice}}$    & - & $5.04 \times 10^{-8}$ \\
C$^+$      & $1.00 \times 10^{-7}$ & - & OCN$_{\rm ice}$       & - & $2.45 \times 10^{-8}$ \\
HCN        & $2.00 \times 10^{-8}$ & - & HNO$_{\rm ice}$       & - & $2.07 \times 10^{-8}$ \\
H$_3^+$    & $1.00 \times 10^{-8}$ & - & C$_2$H$_{\rm ice}$    & - & $9.81 \times 10^{-9}$ \\
C$_2$H     & $8.00 \times 10^{-9}$ & - & S$_{\rm ice}$         & - & $9.00 \times 10^{-9}$ \\
S$^+$      & $9.00 \times 10^{-9}$ & - & CO$_{\rm ice}$       & - & $1.43 \times 10^{-9}$ \\
Si$^+$     & $1.00 \times 10^{-11}$ & - & OH$_{\rm ice}$        & - & $1.21 \times 10^{-9}$ \\
Mg$^+$     & $1.00 \times 10^{-11}$ & - & O$_{\rm ice}$        & - & $1.17 \times 10^{-9}$ \\
Fe$^+$     & $1.00 \times 10^{-11}$ & - & NO$_{\rm ice}$       & - & $2.34 \times 10^{-10}$ \\
N          & - & $1.86 \times 10^{-5}$ & H$_2$CN$_{\rm ice}$   & - & $2.05 \times 10^{-10}$ \\
PAH-H      & - & $1.60 \times 10^{-7}$ & HCO$_{\rm ice}$       & - & $1.48 \times 10^{-10}$ \\
PAH$^+$    & - & $9.61 \times 10^{-8}$ & NH$_{2{\rm ,ice}}$    & - & $1.03 \times 10^{-11}$ \\
NH$_4^+$   & - & $3.20 \times 10^{-8}$ & C$_2$H$_{2{\rm ,ice}}$ & - & $7.28 \times 10^{-12}$ \\
HCO$^+$    & - & $9.00 \times 10^{-9}$ & Si$_{\rm ice}$        & - & $5.87 \times 10^{-12}$ \\
NH$_2$     & - & $9.04 \times 10^{-11}$ & C$_2$N$_{\rm ice}$   & - & $5.02 \times 10^{-12}$ \\
PAH$^-$    & - & $8.95 \times 10^{-11}$ & SiO$_{\rm ice}$      & - & $4.13 \times 10^{-12}$ \\
H$_2$NC$^+$ & - & $2.23 \times 10^{-11}$ & H$_2$CO$_{\rm ice}$ & - & $3.94 \times 10^{-12}$ \\
N$_2$H$^+$ & - & $1.36 \times 10^{-11}$ & & & \\
Mg         & - & $1.00 \times 10^{-11}$ & & & \\
Fe         & - & $1.00 \times 10^{-11}$ & & & \\
S          & - & $3.45 \times 10^{-12}$ & & & \\
H          & - & $2.19 \times 10^{-12}$ & & & \\
\hline                  
\end{tabular}
\end{table*}

\section{Abundance evolution benchmark examples}

\begin{figure*}
\centering
\includegraphics[clip=,width=0.95\linewidth]{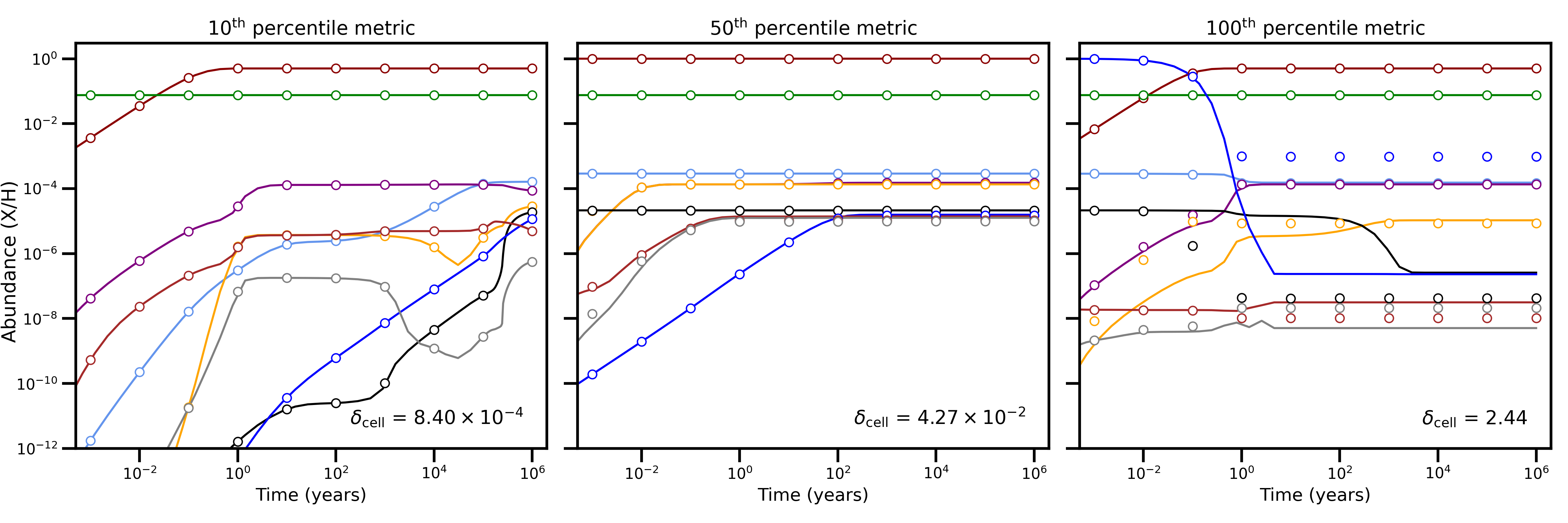}
\caption{Example \textsc{dali} and \textsc{simba} abundance evolutions representative of $10^{\text{th}}$, $50^{\text{th}}$ and $100^{\text{th}}$ percentile metrics. These cells are located in the disk at $(60.0, 4.9)\;\text{au}$, $(120.8, 35.6)\;\text{au}$ and $(0.44, 0.02)\;\text{au}$ respectively. Solid lines represent \textsc{simba} evolution while dots show the \textsc{dali} abundances. For each cell, the 10 most abundant species are displayed.}
\label{fig_benchmark_comparison}
\end{figure*}

To illustrate how our metric captures the level of agreement between \textsc{simba} and \textsc{dali}, Figure \ref{fig_benchmark_comparison} presents examples of abundances over time, with \textsc{simba} compared directly to \textsc{dali}. We show three cases to highlight different levels of agreement: strong (10th percentile, left panel), typical (median, center panel), and poorest (100th percentile, left panel).

\begin{figure}
\centering
\includegraphics[clip=,width=1.0\linewidth]{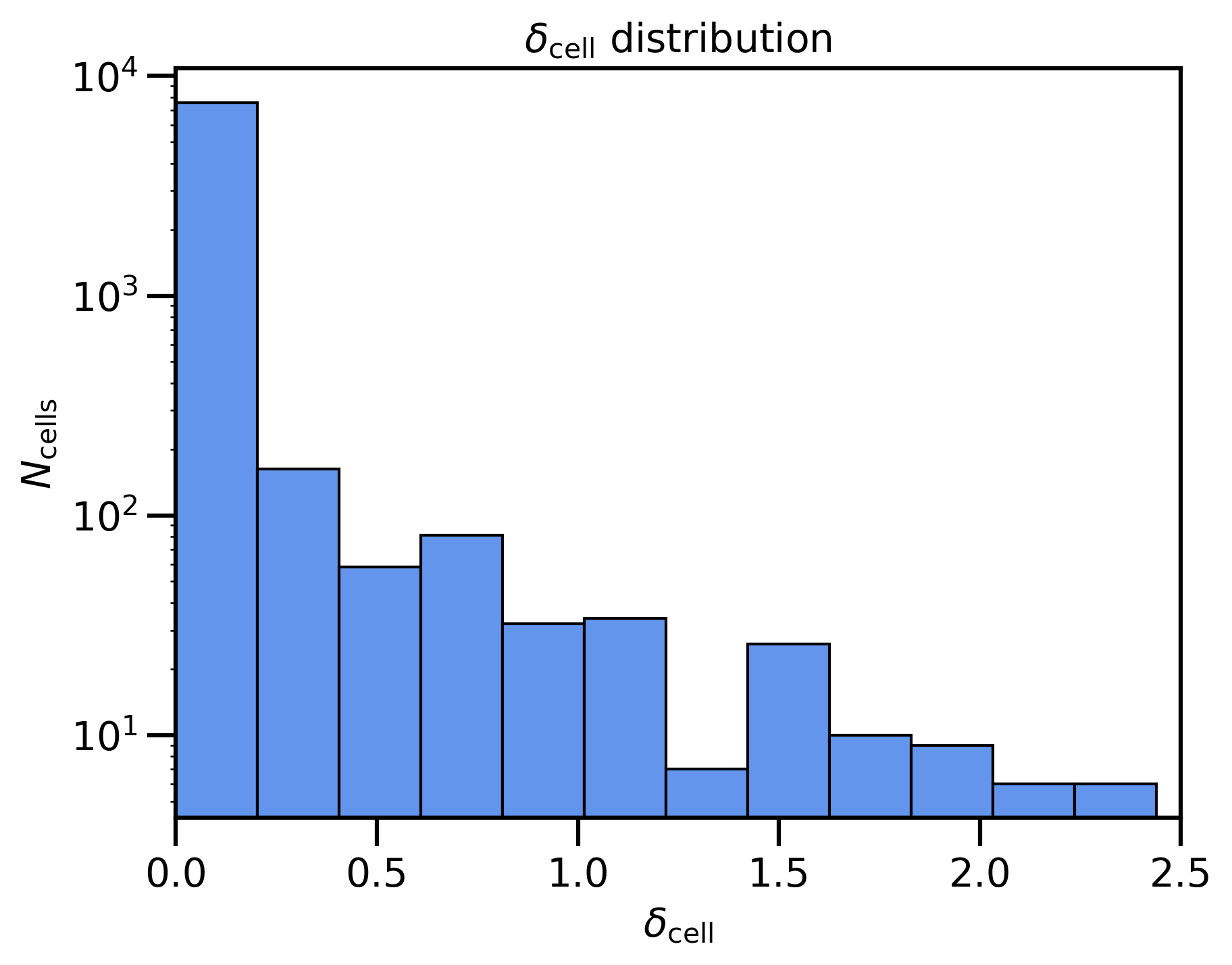}
\caption{Distribution of distance metrics for \textsc{simba} benchmarking across all cells with successful integrations.}
\label{fig_metric_distribution}
\end{figure}

Figure \ref{fig_metric_distribution} displays the distribution of metric values across all instances of successfully integration. The distribution is strongly skewed toward lower values, indicating that \textsc{simba} reproduces the majority of the parameter space well (note the logarithmic scale on the y-axis).

\bsp	
\label{lastpage}
\end{document}